\documentstyle[preprint,eqsecnum,aps,epsf]{revtex}
\newif\iftightenlines\tightenlinesfalse
\tightenlines\tightenlinestrue

\def\eslt{\not\!\!{E_T}}
\def\to{\rightarrow}

\def\te{\tilde e}

\def\tu{\tilde u}
\def\ts{\tilde s}
\def\tb{\tilde b}

\def\td{\tilde d}

\def\tst{\tilde t}
\def\ttau{\tilde \tau}

\def\tg{\tilde g}
\def\tnu{\tilde\nu}
\def\tell{\tilde\ell}

\def\tw{\widetilde\chi^\pm}

\def\tz{\widetilde\chi^0}
\begin{document}
\draft
\preprint{\vbox{\baselineskip=14pt%
   \rightline{FSU-HEP-990702}
   \rightline{UCCHEP/1-99}
   \rightline{IFIC/99-48}
   \rightline{FTUV/99-46}
   \rightline{UH-511-937-99}
}}
\title{SPARTICLE MASS SPECTRA FROM \\ 
SO(10) GRAND UNIFIED MODELS\\
WITH YUKAWA COUPLING UNIFICATION}
\author{Howard Baer$^1$, Marco A. D\'\i az$^{1,2}$, Javier Ferrandis$^{1,3}$
and Xerxes Tata$^4$}
\address{
$^1$Department of Physics,
Florida State University,
Tallahassee, FL 32306 USA
}
\address{
$^2$Facultad de F\'\i sica, Universidad Cat\'olica de Chile,
Av. Vicu\~na Mackenna 4860, Santiago, Chile
}
\address{
$^3$Departament de F\'\i sica Te\`orica,
Universitat de Val\`encia,
Spain
}
\address{
$^4$Department of Physics and Astronomy,
University of Hawaii,
Honolulu, HI 96822, USA
}
\date{\today}
\maketitle
\begin{abstract}

We examine the spectrum of superparticles obtained from the minimal
$SO(10)$ grand unified model, where it is assumed the
gauge symmetry breaking yields the Minimal Supersymmetric Standard Model (MSSM)
as the effective theory at $M_{GUT}\sim 2\times 10^{16}$ GeV.
In this model, unification of Yukawa couplings implies a value of
$\tan\beta\sim 45-55$. 
At such high values of $\tan\beta$, 
assuming universality of scalar masses, the usual mechanism
of radiative electroweak symmetry breaking breaks down.
We show that a set of weak scale sparticle 
masses consistent with radiative electroweak symmetry breaking 
can be generated by imposing non-universal GUT scale scalar masses
consistent with universality within $SO(10)$ plus extra $D$-term
contributions associated with the reduction in rank of the gauge symmetry
group when $SO(10)$ spontaneously breaks to 
$SU(3)\times SU(2)\times U(1)$. We comment upon
the consequences of the sparticle mass spectrum for collider
searches for supersymmetry. One implication of $SO(10)$ unification is that
the light bottom squark can be by far the lightest of the squarks. 
This motivates a dedicated search for bottom squark pair production at
$p\bar{p}$ and $e^+e^-$ colliders.

\end{abstract}

\medskip



Unification of the Standard Model (SM) of strong, weak and 
electromagnetic interactions within a single Lie group such as 
$SU(5)$ or $SO(10)$ has a long history and many attractive 
features\cite{review}. 
$SU(5)$ is the smallest grand unifying group, and predicts the quantization of
electric charge, the unification of gauge couplings and the 
unification of bottom and tau Yukawa couplings at scales of 
$Q=M_{GUT}\simeq 10^{15}$ GeV\cite{su5}. 
The $SO(10)$ theory incorporates all the matter 
fields of the SM into the 16-dimensional spinor representation,
$\psi_{16}$, of
$SO(10)$\cite{so10}. 
In minimal $SO(10)$, not only the gauge couplings but {\it all} the Yukawa
couplings (within a generation) are unified at $Q=M_{GUT}$. 
If the right-handed neutrino field present in $\psi_{16}$
acquires a large Majorana mass, it decouples from the theory,
and a small neutrino mass is induced  via the see-saw 
mechanism\cite{seesaw}.

The supersymmetric version of this model, with supersymmetry (SUSY) softly
broken at a scale $\alt 1$~TeV naturally stabilizes the hierarchy
between the weak scale and the grand-unification scale.
Supersymmetry also raises the
unification scale to $M_{GUT}\simeq 2\times 10^{16}$ GeV, which helps reduce
the rate for proton decay to below the level of experimental bounds.
In addition, the introduction of supersymmetry
with soft SUSY breaking (SSB) masses of order the weak scale allows for the 
near unification of gauge coupling constants\cite{unif}.
In supergravity-based models, it is usually assumed that all scalar masses
receive a common mass $m_Q=m_U=m_D=m_L=m_E=m_{H_u}=m_{H_d}\equiv m_0$ at 
$M_{GUT}$, while all gauginos receive
a common mass $m_{1/2}$ and all trilinear SSB terms unify to $A_0$.
The SSB masses and couplings are then evolved via renormalization 
group equations (RGEs) from $M_{GUT}$ to $Q\sim M_{weak}$. The $m_{H_u}^2$
term is driven to negative values, which results in radiative breaking
of electroweak symmetry, provided the top quark mass is large 
({\it e.g.} 175 GeV).

In addition to the matter superfield $\hat{\psi}_{16}$, the minimal $SO(10)$
model includes a {\bf 10} dimensional Higgs superfield $\hat{\phi}_{10}$ that
decomposes into a ${\bf 5 + \bar{5}}$ representation of $SU(5)$, and
includes the two Higgs superfields ($\hat{H}_u$ and $\hat{H}_d$) of the Minimal
Supersymmetric Standard Model (MSSM).
The superpotential includes the term
\begin{eqnarray*}
W\ni \lambda\hat{\psi}^T\hat{\psi}\hat{\phi} +\cdots
\end{eqnarray*}
responsible for quark and lepton masses, with
$\lambda$ the single Yukawa coupling in the low energy theory.
The dots represent terms including for instance higher dimensional
Higgs representations and interactions responsible for the breaking
of $SO(10)$.

The mass spectrum of SUSY particles in minimal supersymmetric $SO(10)$ 
constrained by radiative electroweak symmetry breaking has been 
studied previously in a number of 
papers\cite{op1,barger,copw,shafi,cw,rsh,mn,hemp,strumia,op2,mop}. 
Unification of bottom, tau and top Yukawa couplings was found to occur at
very large values of the parameter $\tan\beta\sim 50-60$, and specific
spectra were generated for values of $m_t\sim 190$ GeV\cite{barger}.
Assuming universality of
soft SUSY breaking masses at $M_{GUT}$, it was found\cite{copw,cw} 
that Yukawa unification consistent with radiative electroweak
symmetry breaking could also occur for $m_t<170$ GeV as long as
$m_{1/2} \agt 300$ GeV. This generally leads to sparticle 
masses far beyond the reach of the CERN LEP2 or Fermilab Tevatron
$p\bar{p}$ colliders. 
For values of $m_t\simeq 175$ GeV, solutions including 
radiative electroweak breaking were very difficult to achieve.
In Ref. \cite{op2}, the SUSY particle mass spectrum
was investigated with {\it non-universal} SSB masses. Various solutions
were found, but the non-universality in general broke the $SO(10)$
symmetry. In Ref. \cite{mop}, it was argued that $SO(10)$ $D$-term
contributions to scalar masses had the correct form to allow for
successful radiative electroweak symmetry breaking and the computation
of weak scale SUSY particle masses.

In this report, we explicitly calculate the sparticle mass spectrum for 
$SO(10)$ SUSY GUT models, taking the pole mass $m_t=175$ GeV. 
We make the following assumptions.
We assume the structure of minimal SUSY $SO(10)$ above the scale
$Q=M_{GUT}$. We assume that SUSY $SO(10)$ directly breaks to the MSSM 
at $M_{GUT}$. Accordingly, there exist independent masses 
$m_{16}$ and $m_{10}$ for 
the matter and Higgs scalar fields. In the
breakdown of $SO(10)$ to $SU(3)_C\times SU(2)_L\times U(1)_Y$, 
additional $D$-term contributions (parametrized by $M_D^2$ which can be
either positive or negative) to the SSB scalar masses arise\cite{dterms}:
\begin{eqnarray*}
m_Q^2=m_E^2=m_U^2=m_{16}^2+M_D^2 \\
m_D^2=m_L^2=m_{16}^2-3M_D^2 \\
m_{H_{u,d}}^2=m_{10}^2\mp 2M_D^2 . \\
\end{eqnarray*}
Thus, the model is characterized by the following free parameters:
\begin{eqnarray*}
m_{16},\ m_{10},\ M_D^2,\ m_{1/2},\ A_0,\ sign(\mu ).
\end{eqnarray*}
The value of $\tan\beta$ will be restricted by the requirement of
Yukawa coupling unification, and so is {\it not} a free parameter.

Our procedure is as follows. We generate random samples of model
parameters
\begin{eqnarray*}
0&<&m_{16}<1500\ {\rm GeV},\\
0&<&m_{10}<1500\ {\rm GeV},\\
0&<&m_{1/2}<500\ {\rm GeV},\\
-500^2&<&M_D^2<+500^2\ {\rm GeV}^2,\\
45&<&\tan\beta <55, \\
-3000&<& A_0<3000\ {\rm GeV}\ {\rm and} \\
\mu &>&0\ {\rm or}\ \mu <0 .
\end{eqnarray*}
We then calculate the non-universal scalar masses according to
formulae given above, and enter the parameters into the 
computer program ISASUGRA. 
ISASUGRA is a part of the ISAJET package\cite{isajet}
which calculates an iterative solution to the 26 coupled 
RGEs of the MSSM. 

To calculate the values of the Yukawa
couplings at scale $Q=M_Z$, we begin with the pole masses $m_b=4.9$ GeV
and $m_\tau =1.784$ GeV. 
We calculate the corresponding running masses
in the $\overline{MS}$ scheme, and evolve $m_b$ and $m_\tau$ up to 
$M_Z$ using 2-loop SM RGEs. 
At $Q=M_Z$, we include the SUSY loop corrections to $m_b$ and
$m_\tau$ using the approximate formulae of Pierce {\it et al.}\cite{pierce}.
A similar procedure is used to calculate the top quark Yukawa coupling 
at scale $Q=m_t$.

Starting with the three gauge couplings and $t$, $b$ and $\tau$ Yukawa
couplings of the MSSM at scale $Q=M_Z$ (or $m_t$), ISASUGRA evolves the
various couplings up in energy until the scale where $g_1=g_2$, which is
identified as $M_{GUT}$, is reached. The $GUT$ scale boundary conditions
are imposed, and the full set of 26 RGE's for gauge couplings, Yukawa
couplings and relevant SSB masses are evolved down to $Q\sim M_{weak}$,
where the renormalization group improved one-loop effective potential is
minimized at an optimized scale choice $Q=\sqrt{m_{\tst_L}m_{\tst_R}}$
and radiative electroweak symmetry breaking is imposed. Using the new
spectrum, the full set of SSB masses and couplings are evolved back up
to $M_{GUT}$ including weak scale sparticle threshold corrections to
gauge couplings. The process is repeated iteratively until a stable
solution within tolerances is achieved. We accept only solutions for
which the Yukawa couplings $\lambda_t$, $\lambda_b$ and $\lambda_\tau$
unify to within 5\%. This constraint effectively fixes the value of
$\tan\beta$ typically to $\sim 48$.  Yukawa unified solutions are found
only for values of $\mu <0$.  We also require the lightest SUSY particle
to be the lightest neutralino, and that electroweak symmetry
is successfully broken radiatively.

We show in Fig.~\ref{pspace} the regions of model parameter space for
which a SUSY mass spectrum can be calculated consistent with the above
constraints.  In Fig.~\ref{pspace}{\it a}, we show the plane of $m_{10}\
vs.\ m_{16}$.  Each dot represents a point for which a solution was
obtained. Points denoted by a cross are valid solutions, but with
sparticle or Higgs masses below existing limits from LEP2. We require
$m_{\ttau_1}>73$ GeV, $m_{\tw_1}>95$ GeV and $m_h>85.2$
GeV\cite{grivaz}.  From the distribution of points, we see that regions
of model parameter space with $m_{16}<m_{10}$ are preferred, although
for very large values of $m_{16}$, a few solutions are obtained for
$m_{10}>m_{16}$.  In Fig. \ref{pspace}{\it b}, we plot the $M_D\ vs.\
m_{16}$ parameter plane. In this frame, $M_D$ actually stands for
$sign(M_D^2)\times \sqrt{|M_D^2|}$. No solutions were obtained for
$M_D^2<0$, and in fact no solutions were obtained for $M_D^2=0$: this
illustrates that non-zero $D$-term contributions to scalar masses are
crucial for a valid sparticle mass spectrum in minimal $SO(10)$.  The
requirement of positive definite $D$-term contributions to scalar masses
will leave, as we shall see, a distinctive imprint on the SUSY particle
mass spectrum\cite{dterms}. From the $m_{1/2}\ vs.\ m_{16}$
plane in Fig. \ref{pspace}{\it c}, it can be seen that $m_{16}$ is
typically larger than $m_{1/2}$; otherwise $\ttau_1$ becomes the
lightest SUSY particle, in violation of cosmological limits on charged
relic particles. Finally, in Fig. \ref{pspace}{\it d}, we show the range
of $\tan\beta$ values for which solutions were generated versus the
parameter $m_{1/2}$. We see that $46<\tan\beta <52$, with the slightly
higher values of $\tan\beta$ being preferred when $m_{1/2}$ is large.
The bounds on $\tan\beta$ are weakened if $\tau-b-t$ Yukawa unification
is relaxed to more than 5\%.

In Fig.~\ref{masses}, we show the range of selected sparticle and Higgs
boson masses that are generated within minimal $SO(10)$ with Yukawa coupling
unification. In frame {\it a}), we see that the light Higgs boson $h$
has mass generally bounded by $m_h< 125$ GeV. This range of light Higgs boson
masses may well be accessible to Fermilab Tevatron Higgs boson 
searches\cite{conway}. 
Values of $m_h\alt 110$ GeV are associated with cases where $m_A$ becomes
comparable to or smaller than $M_Z$.
In frame {\it b}), we plot solutions in the $\mu\ vs.\ M_2$ plane,
where $M_2$ is the $SU(2)$ gaugino mass.
Many solutions with $|\mu | < M_2$ exist, which generally implies that
the lighter charginos and neutralinos have substantial higgsino 
components. The solutions with large $|\mu |$ and $M_2 \sim 100$~GeV
all correspond to values of $m_{16}>1300$~GeV.
In frame {\it c}), the bottom squark mass
$m_{\tb_1}$ is plotted versus $m_{\tu_R}$. We see that although
$m_{\tu_R}$ can be only as light as $\sim 700$ GeV, the $\tb_1$ mass
can be as low as $\sim 150$ GeV. 
The bottom 
squark (mainly $b_R$) is generically much lighter 
than other squarks, because of the
$D$-term contribution to $m_D$ at $Q=M_{GUT}$ as well as $b$-Yukawa
coupling effects which are significant for large values of $\tan\beta$.
Finally, in frame {\it d}), we show the lightest tau 
slepton mass versus the light chargino mass.
In $SO(10)$, the stau is the lightest of the sleptons, but as can be seen, 
solutions with $m_{\ttau_1}<200$ GeV are very difficult to generate,
and almost always, $m_{\ttau_1}>m_{\tw_1}$, so that two body decays such
as $\tz_2\to\ttau_1\tau$ or $\tw_1\to \ttau_1\nu$ almost never occur.
A feature of minimal $SO(10)$ is that the light stau may
contains a large
left stau component, whereas in models with universality, the light stau
is dominantly a right slepton. This could have an impact on the
efficiency of detecting daughter tau leptons via their hadronic decay.

In Table I, we show sample weak scale sparticle and Higgs boson masses
for five $SO(10)$ solutions with unified Yukawa couplings. It is
possible to find solutions with sparticle masses potentially accessible
to both LEP2 and Fermilab Tevatron searches, in contrast to previous
studies assuming universality of scalar masses at $M_{GUT}$.  For case
1, we take ($m_{16},\ m_{10},\ M_D,\ m_{1/2}, \ A_0) = (405.8,\ 680.3,\
96.8,\ 427.2,\ 596)$ GeV. This solution requires $\tan\beta =51.3$ to
unify the Yukawa couplings. The evolution of gauge and Yukawa couplings
for this case is shown in Fig. \ref{evolve}{\it a}.  In our program, we
do not require the $SU(3)$ gauge coupling to exactly unify with the
$SU(2)$ and $U(1)$ gauge couplings, but rather attribute the near miss
to unknown high scale physics. The Yukawa couplings
diverge from their unification point and evolve to $M_{weak}$, with a
kink in the curves coming from weak scale threshold effects.  In
Fig. \ref{evolve}{\it b}, we show the evolution of SSB Higgs boson
masses and third generation SSB masses.  We actually plot
$sign(m_H^2)\times \sqrt{|m_H^2|}$.  In this case, both Higgs squared
masses evolve to negative values, signaling the onset of radiative
electroweak symmetry breaking. The $GUT$ scale non-universality due to
$D$-term contributions is evident.  It usually results in left SSB
slepton masses being close to or lighter than right slepton masses,
and right sbottom masses lighter than the other squark masses.

The final weak scale sparticle masses are listed in Table I. For case 1,
none of the sparticle or Higgs bosons are accessible to LEP2, while one
or more of the Higgs bosons may be accessible to the Fermilab Tevatron
running at maximal luminosity.  An $e^+e^-$ collider operating at
$\sqrt{s}=500$ GeV would find not only the various MSSM Higgs bosons,
but also charginos and neutralinos (with substantial higgsino
components) and light tau sleptons (in this case $\ttau_1 \sim \ttau_R$).

The second case study point shown has very large values of
$m_{16}$ and $m_{10}$, leading in general to a spectrum with very 
heavy scalars. The exception in this case is that the $\tb_1$ mass
is only 140 GeV, and is directly accessible to 
Fermilab Tevatron collider searches, according to Ref. \cite{bmt}. 

In this case,  
$\tb_1\to b\tz_1$ with a branching fraction 
of  $\sim 80\%$, so that $\tb_1\bar{\tb_1}$ production
would be visible in $b\bar{b}+\eslt$ events.
The gluino is also relatively light and decays via $\tg \to b\tb_1$: it would
be interesting to examine whether the improved $b$ tagging at Tevatron
upgrades would allow its detection in the multi-$b$ plus $\eslt$ channel.
The light Higgs boson might be accessible to high luminosity Tevatron
experiments, but the charginos and
neutralinos would be difficult to see via the trilepton channel
since $\tz_2$ dominantly
decays to $b\bar{b}\tz_1$ and the $\tz_2\to e\bar{e}\tz_1$ 
branching fraction is only $0.8$\%.
A Higgs signal of $b\bar{b}\ell +\eslt$ events 
from $Wh$ production would contain substantial contamination from
$\tw_1\tz_2$ events, which give rise to the same event topology.
Contrary to models with universality, the left selectron (smuon)
is significantly lighter than the right selectron (smuon). 
This distinctive feature of the $SO(10)$ model would be
difficult to discern as the sleptons are very heavy.
In the squark sector, the $\td_R$ and $\ts_R$ 
are the lightest of the first two generations of squarks, owing 
to the $D$-terms. 

In case 3, again the bulk of the scalars are quite heavy, and well
beyond the reach of LEP2 or the Tevatron. Again the exception is the light 
bottom squark. In this case, however, $\tb_1$ decays with a 24\% (8\%) 
branching fraction to $b\tz_2$ ($b\tz_3$), so the event signatures will be
more complicated. Since $\tz_2$ and $\tz_3$ decay with a large rate
to $b\bar{b}\tz_1$, some of the $\tb_1\bar{\tb_1}$ events will contain 
final states with up to six $b$-jets plus $\eslt$!
If clean trilepton signatures are detected\cite{bdpqt}, 
they will contain a mixture
of events from both $\tw_1\tz_2$ and $\tw_1\tz_3$ production.

In case 4, all strongly interacting sparticles including the bottom squark 
are quite heavy and accessible only at the LHC.
However, the various sleptons and sneutrinos are within reach of an
$e^+e^-$ linear collider operating at $\sqrt{s}\simeq 1000$ GeV.
In this case, a very mixed $\ttau_1$
whose composition may be measurable at a Linear
Collider \cite{nft} may serve to distinguish this framework from models
with universal soft masses. 
Moreover, $\tnu_{eL}$
and the $\te_L$ are measurably lighter \cite{munroe} than $\te_R$,
again in contrast with 
expectations in models with universality.
Note also that
$|\mu |<M_2$, so that the light charginos and neutralinos have
substantial Higgsino components, 
and further that there is only a small mass gap between
$m_{\tz_2}$ or $m_{\tw_1}$ and $m_{\tz_1}$, so that -ino decay products
will be soft.

Finally, in case 5, again the light charginos and neutralinos are
higgsino-like, and will be challenging to detect at the Fermilab
Tevatron collider. This spectrum is characterized by a very light Higgs boson
spectrum, and in fact 32\% of top quark decays are to charged Higgs bosons.
Indeed, both the light and pseudoscalar Higgs boson are at the edge of 
detectability at the LEP2 collider.

We have demonstrated
that the inclusion of $D$-terms can lead to radiative electroweak
symmetry breaking even in models with Yukawa coupling unification. 
As shown in Fig.~\ref{pspace}{\it b}, we
are unable to find corresponding solutions for models with scalar mass
universality
for the ranges of parameters studied here. 
We have not attempted to do an analysis of the phenomenological
implications of the model. In a follow-up report, we will present
results of calculations for the neutralino relic density, $b\to s\gamma$
decay rate, direct dark matter detection rate, and prospects for
collider searches\cite{bbdfmqt}. 
Parts of the parameter space as
well as some of the case studies may well be excluded by experimental
constraints. For instance, our
preliminary results indicate that the predicted value for
the decay $b \rightarrow s\gamma$ exceeds the experimental upper limit
by a factor $\sim 2-4$ if $m_{1/2} \sim 200-500$~GeV. This is well-known
to be a problem common to models with large $\tan\beta$ and 
$\mu <0$\cite{bbct}, the region of parameter space where Yukawa
couplings unify.

Despite this phenomenological problem, we find it encouraging that it is
possible to construct a calculable framework with gauge and Yukawa
coupling unification. We can imagine other physics that may make it
possible to circumvent experimental limits such as those from
$b\to s\gamma$. For example, it has
been pointed out~\cite{bagger} that if the right-handed neutrino mass is
significantly below the GUT scale and if $m_{10}^2 = 2m_{16}^2$, third
generation scalars would be radiatively driven to much lower masses than
other matter scalars, and further, that when $D$-terms are included,
radiative electroweak symmetry breaking is still possible \cite{bmt2}. In
this case, since the degeneracy between squarks is badly broken, it is
possible that gluino-mediated contributions to $b \rightarrow s\gamma$
(which are generally thought to be small) may be significant. Whether
these are large enough (and of the correct sign) to cancel the
chargino-mediated amplitudes remains to be
investigated. Alternatively, one might imagine that the usual
computation of $b\rightarrow s\gamma$
amplitudes may be altered by $CP$ violating phases between various
chargino amplitudes\cite{bk}, or by large Yukawa coupling radiative 
corrections\cite{wagner}.

{\it Summary:} 
We have shown that explicit evaluation of 
sparticle mass spectra 
is possible in the minimal SUSY $SO(10)$ model with Yukawa coupling 
unification and radiative electroweak symmetry breaking, 
by including non-universal SSB masses at $Q=M_{GUT}$ which are
in accord with $SO(10)$ breaking to the gauge group of the MSSM. The 
resulting spectra reflect the influence of the $D$-term contributions 
to scalar masses. 
Characteristic features of the model can include a light sbottom,
a $\ttau_1$ which is mainly $\ttau_L$, $m_{\tell_L} < m_{\tell_R}$
and lighter charginos and
neutralinos with substantial (sometimes even dominant) Higgsino components.


%
\acknowledgments
We thank Damien Pierce and Konstantin Matchev for discussions.
This research was supported in part by the U.~S. Department of Energy
under contract numbers DE-FG02-97ER41022 and DE-FG03-94ER40833.
J.F. was supported by a Spanish MEC FPI fellowship, a travel grant
from Generalitat Valenciana, by DGICYT under grants PB95-1077
and by the TMR network grant ERBFMRXCT960090 of the E.U.
%

%

\newpage
%
%

\iftightenlines\else\newpage\fi
\iftightenlines\global\firstfigfalse\fi
\def\dofig#1#2{\epsfxsize=#1\centerline{\epsfbox{#2}}}

\begin{table}
\begin{center}
\caption{Weak scale sparticle masses and parameters (GeV) for five $SO(10)$
case studies.}
\bigskip
\begin{tabular}{lccccc}
\hline
parameter & case 1 & case 2 & case 3 & case 4 & case 5 \\
\hline

$m_{16}$ & 405.8   & 1240.0 & 1022.0 & 414.8 & 629.8 \\
$m_{10}$ & 680.3   & 1414.0 & 1315.0 & 735.7 & 836.2\\
$M_D   $ & 96.8 & 410.6 & 329.8 & 171.9 & 135.6 \\
$m_{1/2}$ & 427.2  & 136.5 & 232.0 & 449.1 & 348.8 \\
$A_0$ & 596.0 & -1100.0  & -1350.0 & 576.7 & -186.5 \\
$\tan\beta$ & 51.3 & 47.0 & 48.6 & 51.3 & 52.1 \\
$m_{\tg}$ & 1021.3 & 409.9 & 631.5 & 1069.3 & 864.8 \\
$m_{\tu_L}$ & 983.7 & 1337.8 & 1178.5 & 1033.7 & 974.4 \\
$m_{\td_R}$ & 925.4 & 1057.6 & 970.1 & 934.9 & 910.8 \\
$m_{\tst_1}$& 718.4 & 737.9  & 512.3 & 754.5 & 618.7 \\
$m_{\tb_1}$ & 735.6 & 140.6 & 187.1 & 721.7 & 636.8 \\
$m_{\tell_L}$ & 478.7 & 1012.8 & 857.8 & 428.3 & 634.6 \\
$m_{\tell_R}$ & 452.9 & 1321.1 & 1088.9 & 489.6 & 662.5 \\
$m_{\tnu_{e}}$ & 472.0 & 1009.7 & 854.1 & 420.7 & 629.5 \\
$m_{\ttau_1}$ & 233.2  & 790.1 & 623.6 & 272.6 & 427.8 \\
$m_{\tnu_{\tau}}$ & 386.0 & 787.4 & 619.5 & 314.6 & 519.1 \\
$m_{\tw_1}$ & 159.5 & 110.5 & 122.9 & 177.5 & 106.3 \\
$m_{\tz_2}$ & 166.7 & 110.3 & 131.6 & 195.1 & 126.1 \\
$m_{\tz_1}$ & 129.0 & 56.8  & 84.0 & 152.3 & 87.5 \\
$m_h      $ & 113.7 & 115.5 & 118.8 & 116.4 & 93.7 \\
$m_A      $ & 115.8 & 645.0 & 479.9 & 277.9 & 93.9 \\
$m_{H^+}  $ & 152.2 & 652.3 & 490.2 & 295.1 & 137.1 \\
$\mu      $ & -157.2 & -329.8 & -150.5 & -185.5 & -113.9 \\
$\langle\ttau_1 |\ttau_L\rangle$ & 
0.14 & 0.99 & 0.99 & 0.47 & 0.11 \\
\hline
\label{tab:cases}
\end{tabular}
\end{center}
\end{table}

\newpage
%

%
\begin{figure}
\dofig{5in}{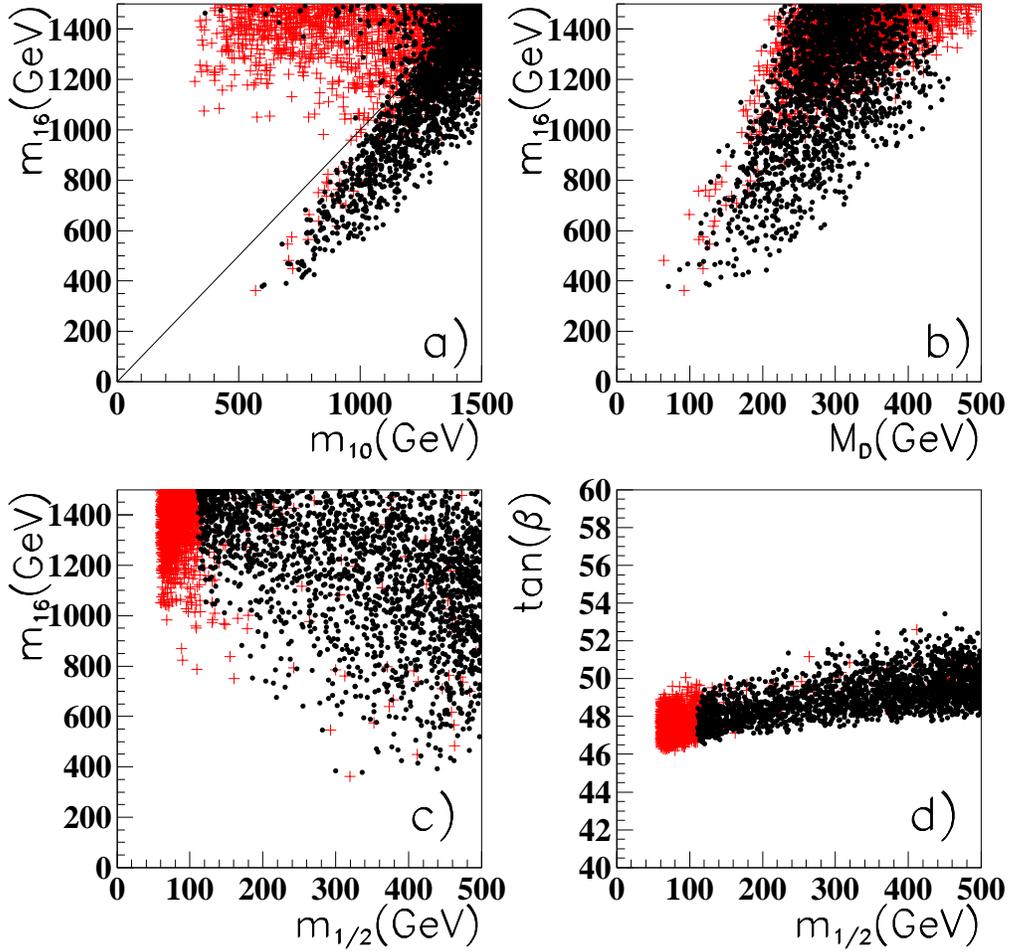}
\caption[]{
Plots of regions of parameter space where valid solutions to 
minimal SUSY $SO(10)$ are obtained, consistent with Yukawa
coupling unification to 5\%, and radiative electroweak symmetry breaking.}
\label{pspace}
\end{figure}
\begin{figure}
\dofig{5in}{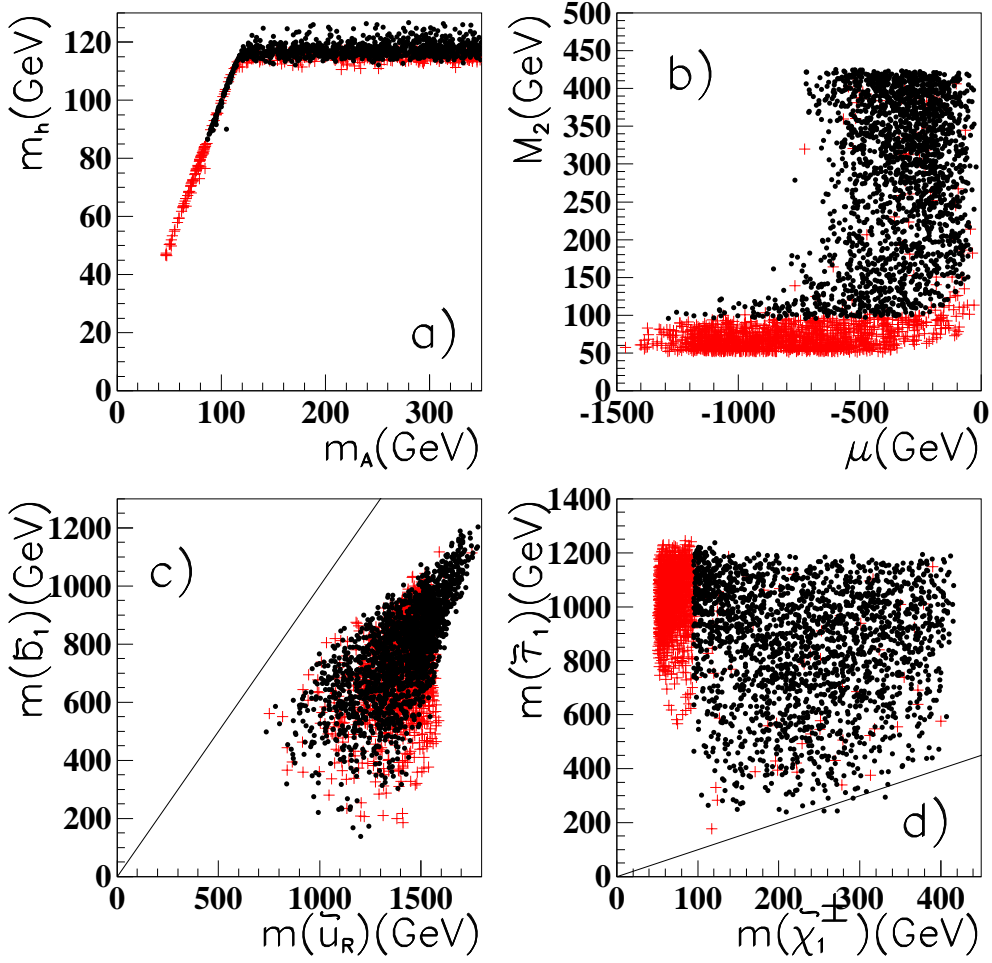}
\caption[]{
The range of selected sparticle masses that are generated in 
minimal SUSY $SO(10)$ models with Yukawa coupling unification and
radiative electroweak symmetry breaking.}
\label{masses}
\end{figure}
\begin{figure}
\dofig{4in}{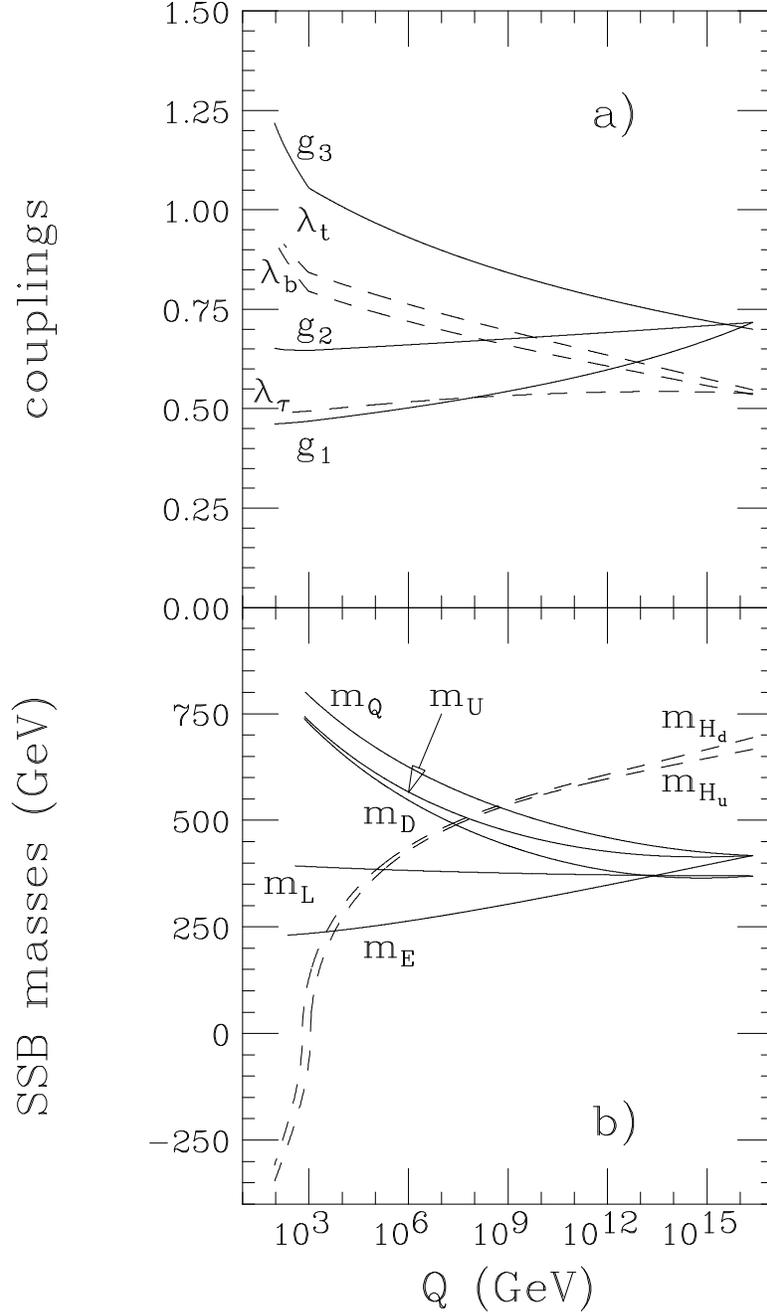}
\caption[]{
For case 1 in Table I, we show {\it a}) the running of both gauge and Yukawa
couplings between $Q=M_{GUT}$ and $Q=M_{weak}$.
In {\it b}), we show the running of SSB Higgs masses (dashed curves)
and third generation SSB masses (solid curves).}
\label{evolve}
\end{figure}

\end{document}